# Contact Resistance Optimization in MoS$_2$ Field-Effect Transistors through Reverse Sputtering-Induced Structural Modifications

Yuan Fa[1,2], Agata Piacentini[1,2], Bart Macco[3], Holger Kalisch[4], Michael Heuken[4,5], Andrei Vescan[4], Zhenxing Wang[1,*], and Max C. Lemme[1,2,*]

[1] AMO GmbH, Advanced Microelectronic Center Aachen, Otto-Blumenthal-Str. 25, 52074 Aachen, Germany

[2] Chair of Electronic Devices, RWTH Aachen University, Otto-Blumenthal-Str. 25, 52074 Aachen, Germany

[3] Department of Applied Physics and Science Education, Eindhoven University of Technology, 5600 MB Eindhoven, The Netherlands

[4] Compound Semiconductor Technology, RWTH Aachen University, 52074 Aachen, Germany

[5] AIXTRON SE, 52134 Herzogenrath, Germany

**Abstract**

Two-dimensional material (2DM)-based field-effect transistors (FETs), such as molybdenum disulfide (MoS$_2$)-FETs, have gained significant attention for their potential for ultra-short channels, thereby extending Moore's law. However, MoS$_2$-FETs are prone to the formation of Schottky barriers at the metal-MoS$_2$ interface, resulting in high contact resistance ($R_c$) and, consequently, reduced transistor currents in the ON-state. Our study explores the modification of MoS$_2$ to induce the formation of conductive 1T-MoS$_2$ at the metal-MoS$_2$ interface via reverse sputtering. MoS$_2$-FETs exposed to optimized reverse sputtering conditions in the contact area show $R_c$ values reduced to less than 50% of their untreated counterparts. This reduction translates into improvements in other electrical characteristics, such as higher ON-state currents. Since reverse sputtering is a standard semiconductor process that enhances the electrical performance of MoS$_2$-FETs, it has great potential for broader application scenarios in 2DM-based microelectronic devices and circuits.

# Introduction

Silicon-based field-effect transistors (FETs), the fundamental components of integrated circuits, are approaching their physical limits at physical gate lengths of approximately 10 nm.[1] New materials may be capable of overcoming these constraints, thus enabling the continuation of miniaturizing integrated circuits. Molybdenum disulfide ($MoS_2$), a semiconducting two-dimensional (2D) transition metal dichalcogenide (TMDC), has emerged as a promising channel material because of its potential for next-generation FETs.[2–11] Its ultra-thin structure enables ideal electrostatic control of the channel potential by the gate electrode, facilitating low OFF-currents ($I_{off}$). In addition, high ON-currents ($I_{on}$) are achievable through multiple-channel stacking.[12,13] Additionally, in contrast to silicon, the layered nature of $MoS_2$ enables it to maintain excellent carrier mobility down to the monolayer regime.[11,14,15]

$MoS_2$ is a polymorphic material and exists in different structural forms, with the most common being the trigonal prismatic coordination (2H) and octahedral symmetry (1T) phases. The 2H phase, which is thermodynamically stable and commonly found in natural $MoS_2$ crystals, is semiconducting. Bulk 2H-$MoS_2$ is a semiconductor with an indirect bandgap of 1.2 eV, which transitions to a direct bandgap of 1.8 eV for monolayer $MoS_2$.[16] Conversely, the 1T phase of $MoS_2$ is metallic, with a conductivity $10^7$ times greater than that of the 2H phase.[17,18] It is metastable and can be chemically synthesized. In addition, the transition from the 2H to 1T phase can be achieved by shifting a sulfur atom within the unit cell from one pyramidal position to another.[19]

$MoS_2$ has become an attractive candidate among TMDs for advanced transistor applications due to its tunable band gap and polymorphism, moderately high carrier mobility[3] even at the monolayer limit, and mechanical flexibility.[20] MoS$_2$-FETs have been proposed for analog and digital electronics because of their large on/off-current ratios of up to $10^{10}$, mobility of approximately 200 cm²·V⁻¹·s⁻¹, and a subthreshold swing approaching the theoretical limit of 60 mV/dec at room temperature.[9] Despite these excellent properties, achieving low contact resistances ($R_c$) remains challenging because a Schottky barrier forms at the MoS$_2$-metal interface. Unlike conventional transistors, where low $R_c$ can be achieved by high (degenerate) doping of the semiconductor, this approach is not applicable in MoS$_2$-based FETs. The large Schottky barrier height (SBH) restricts electron injection and degrades performance.[21,22]

Several approaches have been devoted to lowering the SBH and reducing $R_c$ in MoS$_2$-FETs. Promisingly, zero SBH and ohmic contacts to monolayer TMDCs have been achieved with the semimetals bismuth (Bi) and antimony (Sb),[23–25] although their low abundance may pose challenges for sustainable large-scale production. Contact deposition under ultrahigh vacuum (UHV) conditions has also been shown to reduce $R_c$ to MoS$_2$-FETs, even with complementary metal oxide semiconductor (CMOS) technology-compatible metals.[26] However, UHV deposition may not be cost-efficient for mass production. Recently, low $R_c$ was achieved using heterostructures of semimetallic (CVD-grown) monolayer graphene and MoS$_2$.[6,27] However, fabricating 2D heterostructures increases the complexity of this approach. Kappera *et al.* presented a chemical method to induce a MoS$_2$ phase transition to design heavily doped source and drain contacts to achieve a low $R_c$.[4] However, this method has limitations due to the chemical toxicity involved, and it has not yet been demonstrated on polycrystalline CVD materials required for wafer-scale integrated circuits. Furthermore, MoS$_2$-FETs have been exposed to argon (Ar) plasma to reduce $R_c$ by either inducing a phase transition to 1T-MoS$_2$ at the metal contact interface[7] or by creating sulfur vacancies (V$_S$)[8], although more detailed studies are required to understand the relationship between the phase transition and defect concentration in MoS$_2$.

Reverse sputtering employs Ar ions produced by radio frequency (RF) generated plasma, which are attracted to the substrate by a negative bias in a vacuum chamber. This method is attractive because of its cleanliness, scalability, and controllability, and it can be combined *in situ* with metal sputter deposition, enabling a controlled contact fabrication process. Here, we demonstrate a scalable methodology to reduce $R_c$ in metal-organic chemical-vapor-deposited (MOCVD) MoS$_2$-based FETs through reverse sputtering. We analyzed MoS$_2$ with photoluminescence (PL) and Raman spectroscopy to verify the influence of reverse sputtering at process parameters and conducted X-ray photoelectron spectroscopy (XPS), which indicated the existence of 1T phase MoS$_2$ and S vacancies in the sputter-treated material. We fabricated MoS$_2$-FETs with nickel (Ni) top-contacts (TOP-FETs) to the sputter-induced 1T-MoS$_2$ and pristine 2H-MoS$_2$ as the channel, and control FETs with bottom-contacts (BOT-FETs) where the entire MoS$_2$, i.e. channel and electrode regions, was exposed to the Ar plasma. While the former devices showed improved R$_c$, the latter lost gate modulation. This demonstrates the feasibility of our approach toward scalable contact engineering in MoS$_2$-FETs.

## Results and Discussion

All experiments were carried out on silicon (Si) chips (2 × 2 cm$^2$) covered with 90 nm thermal SiO$_2$. The MoS$_2$ material was grown via metal-organic chemical vapor deposition (MOCVD) on sapphire substrates. Details of the material growth and device fabrication are described in the Methods section.

Atomic force microscopy (AFM) was conducted to measure the layer thickness of the pristine MOCVD MoS$_2$ on the as-grown sapphire wafer (Fig. 1a). The inset height profile shows a MoS$_2$ film height of 2.5 nm, corresponding to approximately four layers if we assume a thickness of monolayer MoS$_2$ of 0.65 nm.[28]

We then prepared four SiO$_2$/Si substrates with MoS$_2$ transferred onto them. One sample was kept in its pristine state without any treatment, whereas the other three samples were subjected to reverse sputtering at different power levels of 25 W, 75 W, and 200 W for 10 s. Afterwards, PL and Raman spectroscopy were conducted. Fig. 1b shows the PL spectra of pristine MoS$_2$ and the sputter-treated MoS$_2$ films. A distinct peak is observed at approximately 1.80 eV in the PL of pristine MoS$_2$, corresponding to the characteristic A exciton and indicating its semiconductor properties.[29] Sputter treatments with 25 W, 75 W, and 200 W resulted in PL quenching, which was stronger for increasing power. The PL quenching can be attributed to an increasing vacancy or defect concentration and a transition from a semiconducting state to a metallic state, namely, the 2H to 1T transition (see also XPS data below).[7,30,31] We also noticed a shift toward higher PL photon energy upon sputter exposure. The PL spectra obtained at 25 W were also gradually quenched as the duration increased (see supporting information (SI) Fig. S1).

Following this initial experiment, the three sputter-treated samples underwent additional 10-second increments of sputter exposure with the same power until a total duration of 50 s was reached. We again used Raman spectroscopy to investigate the MoS$_2$. Fig. 1c-e show the average Raman spectra of the MoS$_2$ in an area of 5 μm × 5 μm obtained at the different sputtering power levels (see details in the Methods section). Each graph shows the spectra taken after 10, 20, 30, 40, and 50 s of sputter exposure and includes a Raman spectrum of pristine MoS$_2$ for reference (purple). The distinct peaks corresponding to the in-plane ($E_{2g}^1$) and out-of-plane ($A_{1g}$) modes are observed at approximately 382.0 cm$^{-1}$ and 406.7 cm$^{-1}$, with a position difference Δ of 24.7 cm$^{-1}$. Additionally, a small new peak labeled ***J*** emerges at ~277 cm$^{-1}$ (more details of the intensity of the ***J*** peak in SI Fig. S2a), suggesting the creation of a distorted structure that includes a 2a$_0$ × a$_0$ basal plane superlattice and distorted octahedral coordination in the MoS$_2$ crystalline structure.[32] The Raman spectrum of the sample at 25 W for 10 s in Fig. 1c shows increased intensities of the $E_{2g}^1$ and $A_{1g}$ peaks, possibly due to alterations in the electronic band structure enhancing the interaction between electrons and $A_{1g}$ phonons.[33] Combined with the PL data of 25 W for 10 s, this suggests a different crystalline structure in MoS$_2$ compared with pristine MoS$_2$ and other reverse sputtering parameters. In addition, the $E_{2g}^1$ and $A_{1g}$ peaks in Fig. 1d-e exhibit a notable redshift, which becomes more pronounced with increasing sputtering duration, with a maximum value of approximately 2 cm$^{-1}$. This indicates weakened in-plane and out-of-plane vibrations of the Mo-S bonds and is a typical signal of a reduced layer number of MoS$_2$,[34,35] meaning that the overall thickness of the multilayer MoS$_2$ in the 5 μm × 5 μm area was slightly reduced by reverse sputtering. Moreover, as the treatment duration and sputtering power increase, the $E_{2g}^1$, $A_{1g}$, and ***J*** peaks exhibit broadening and intensity degradation (more details in SI Fig. S2), culminating in a flat spectrum for the case of 200 W for 50 s in Fig. 1e. This indicates a progressive loss of MoS$_2$ crystalline structure, the emergence of V$_S$, structural disruption, and the potential presence of MoO$_x$.[36,37]

XPS was used to evaluate the effects of reverse sputtering on the stoichiometry and binding configuration. It was performed on the pristine multilayer MoS$_2$ sample and the three samples exposed to 10 s of reverse sputtering with 25, 75, and 200 W of power. The pristine MoS$_2$ sample has an S/Mo ratio close to 2 (Fig. 2a). Reverse sputtering leads to a reduced stoichiometry, with the S/Mo ratio of the 200 W sample decreasing to 1.80, which could be a result of a large number of V$_S$ in the crystalline structure of MoS$_2$. Fig. 2b-c show the Mo3d and S2p spectra, respectively. For pristine MoS$_2$, the binding energies of Mo3d$_{5/2}$ and Mo3d$_{3/2}$ are 230.1 eV and 233.2 eV, and the binding energies of S2p$_{3/2}$ and S2p$_{1/2}$ are 162.9 eV and 164.1 eV, respectively. Qualitatively, reverse sputtering induces broadening of the Mo3d, S2s, and S2p peaks as well as a shift to lower binding energies. The broadening is a sign of increased disorder in the system, whereas the binding energy shift is thought to originate from the 2H to 1T phase transition, both induced by Ar$^+$ ion bombardment, in line with the literature.[7,8] To quantify the phases, the Mo3d$^{4+}$ peak was fitted with two contributions. The 1T contribution was constrained to have a binding energy 0.9 eV higher than the 2H phase contribution.[38] The pristine sample predominantly displays a semiconducting 2H phase, with a minor fitted metallic 1T phase. Moreover, the accurate quantification of oxygen in the MoS$_2$ layers is hindered because the MoS$_2$ layers are placed directly on a 90 nm SiO$_2$ film. However, we observed an additional weak peak at approximately 236 eV in the XPS spectrum of the 200 W 10 s sample, corresponding to the higher oxidation state of Mo$^{+6}$.[39] An increase in the O1s contribution at approximately 531 eV is observed after reverse sputtering (Fig. 2d), which indicates enhanced oxidation of the MoS$_2$ layers in the ambient environment after reverse sputtering and a stronger binding effect of Mo-O.[40] While pristine MoS$_2$ is relatively resilient to oxidation, reverse sputtering effectively creates $V_S$, which, upon ambient exposure, can be more easily oxidized to $MoO_x$. After all the sputter treatments, only a metallic 1T phase is obtained from the fitting procedure in Fig. 2b, implying full-phase conversion. As the 2H to 1T phase transition can be driven by $V_S$,[5,7] this is also in line with the reduced S/Mo ratio, as shown in Fig. 2a. While XPS suggests that 10 s of reverse sputtering at 25 W is sufficient to complete the 2H to 1T phase transition, higher sputtering powers result in additional modification of $MoS_2$: the S/Mo ratio decreases, additional broadening of the S2p peak is observed, and an additional O1s contribution appears. Hence, there seems to be a balance between phase conversion and excessive damage to $MoS_2$ via reverse sputtering.

After the basic material characterization, we investigated the influence of reverse sputtering on the performance of $MoS_2$-FETs with a channel width (W) of 25 μm. We conducted electrical measurements on 4 different samples of TOP-FETs in a vacuum: a pristine sample and three sputter-treated samples treated with 25 W, 75 W, and 200 W for a fixed duration of 10 s; the same parameters used for the XPS analysis. Notably, the channel $MoS_2$ remained in a pristine state in the top-contact configuration as it was protected by a photoresist during the optical lithography process (see Methods), while the $MoS_2$ under the source and drain metal was subjected to reverse sputtering as shown in the schematic and the microscope image in Fig. 3a). More details on the fabrication process are provided in the Methods section and SI Fig. S3. In Fig. 3b, the output characteristics of four TOP-FETs with a channel length ($L_{ch}$) of 4 μm are presented for each sample. Most evidently, the output current ($I_{out}$) of the device subjected to 75 W reverse sputtering is significantly greater by approximately 150 times than that of the pristine TOP-FET, whereas $I_{out}$ returns to the level of the pristine TOP-FET when exposed to 200 W reverse sputtering. Fig. 3c shows the statistical distribution of the two-probe FET mobility ($\mu_{2p}$) for seven FETs with increasing $L_{ch}$ on each sample, which was calculated via equations 1 and 2 from 2-terminal measurements:

$$\mu_{2p} = \frac{L_{ch}}{WC_iV_{SD}}\left(\frac{\partial I_D}{\partial V_{GS}}\right) \quad (1)$$

$$C_i = \frac{\epsilon_{SiO_2}\epsilon_0}{t_{ox}} \quad (2)$$

$C_i$ represents the gate-channel capacitance per unit area. $\epsilon_{SiO2}$ represents the relative permittivity of silicon dioxide, equal to 3.9. $\epsilon_0 = 8.854 \times 10^{-12}$ F·m$^{-1}$ represents the vacuum permittivity, and $t_{ox}$ represents the oxide thickness. In our case, $L_{ch}$ ranges from 4 to 10 μm, W equals 25 μm, and $t_{ox}$ is 90 nm. $\mu_{2p}$ is derived from two-probe measurements, which include a voltage drop across the Schottky barrier between 2D MoS$_2$ and metal contacts.[41] Therefore, $\mu_{2p}$ serves as an indicator of the contact performance of the TOP-FETs, provided that the channel material is of the same quality. By fitting a Gaussian distribution to the histogram data, the mean value and standard deviation of $\mu_{2p}$ for each sample are as follows: 0.48 ± 0.10, 0.52 ± 0.11, 0.81 ± 0.20, and 0.17 ± 0.06 cm$^2$·V$^{-1}$·s$^{-1}$ for the pristine sample, 25 W, 75 W, and 200 W treatments, respectively. An increase in $\mu_{2p}$ for the 75 W sample is observed, nearly doubling compared with that of the pristine sample. While the 25 W sample shows barely any increase in $\mu_{2p}$, the 200 W sample exhibits degradation. Fig. 3d-g show n-type transfer characteristics of seven TOP-FETs with increasing ($L_{ch}$) for a $V_{DS}$ of 1 V for pristine and different treatment powers. The TOP-FETs on the pristine, 25 W, and 75 W samples (Fig. 3d-f) exhibit on/off current ratios of 10$^6$, and the transfer curves on a linear scale clearly reveal an increase in the $I_{on}$ with increasing power. However, the on/off ratio of the 200 W sample degrades to $10^5$, and a decrease in $I_{on}$ can also be observed. These transfer characteristics are compatible with the output characteristics described in Fig. 3b.

Fig. 4a-d displays the Transfer Length Method (TLM) plots for 28 TOP-FETs in 4 groups with $L_{ch}$ ranging from 4 to 10 μm on each sample under an overdrive voltage ($V_{OV}$) of 19 V at a carrier concentration ($n_S$) of approximately $4.3 \times 10^{12}$ cm$^{-2}$. Table 1 presents the corresponding average values of $R_c$ and the sheet resistance ($R_{sh}$) as well as their standard deviations extracted from TLM. The $R_c$ of the sample subjected to 75 W decreased to 413 kΩ·μm from 1126 kΩ·μm for the pristine sample. The underlying reason for the improved contact performance lies in two factors. On the one hand, $V_S$ introduced by reverse sputtering can create highly n-type doped regions in MoS$_2$ crystalline structures. This doping narrows the Schottky barrier, enabling electrons to tunnel through it more easily, thereby reducing $R_c$.[8] On the other hand, $V_S$ promote the phase transition from 2H to 1T, which is structurally achieved by transverse displacement of the S planes. In this case, conductive 1T-MoS$_2$ forms more easily, enhancing the injection of electrons from the Ni into the metallic 1T phase and then from the 1T into the 2T phase.[17,19] Furthermore, the $R_c$ of the 25 W sample increases by approximately 50% compared with that of the pristine sample. The reason may be that the reverse sputtering does not create a sufficient number of $V_S$ with this parameter. In addition, the $R_c$ of the 200 W sample is significantly higher than that of the other samples, which can likely be attributed to substantial oxidation to MoO$_x$ caused by a large number of $V_S$, as supported by the XPS results. The presence of MoO$_x$ in the MoS$_2$ structure distorts the crystalline structure of MoS$_2$, which reduces the conductivity of the 2D material under the contact metal and degrades the electrical performance of the TOP-FETs.[42] The $R_{sh}$ values of the reference sample and the 25 W and 75 W samples remained at the same resistive level, indicating that the pristine channel $MoS_2$ was unaffected by reverse sputtering. However, the $R_{sh}$ of the 200 W sample was exceptionally high. We attribute this to excessive damage to the $MoS_2$ channel during the high-power sputtering process and the subsequent exposure of the highly defective $MoS_2$ channel to the ambient environment, which resulted in the formation of highly resistive $MoO_x$. We also fabricated bottom-contact BOT-FETs as a control experiment, where we deposited the source and drain contacts before transferring the $MoS_2$ films (more details of the device fabrication are provided in the Methods section and SI Figs. S4 and S5a). Here, we exposed the entire $MoS_2$ in the channel and contact areas to the sputter treatment. Electrical measurements conducted in ambient air conditions showed that the total resistance ($R_{tot}$) decreased by several orders of magnitude upon high-power sputtering (SI Fig. S5b). However, apart from the case of 25 W for 10 s, where weak gate control is still observed, the transfer curves of BOT-FETs under 75 W and 200 W exhibit a complete loss of gate control. Additionally, the drain current ($I_D$) increases by more than five orders of magnitude compared to the $I_{off}$ of the pristine BOT-FET (SI Fig. S5c-e). These combined results collectively confirm a transition of the $MoS_2$ to the metallic phase.

## Conclusion

We investigated the potential of reverse sputtering for reducing the contact resistance to FETs with 2D $MoS_2$ channel. We show that the $Ar^+$ bombardment in a reverse sputter process modifies the crystalline structure of multilayer MOCVD $MoS_2$ from pristine 2H-$MoS_2$ to metallic 1T-$MoS_2$ through PL, Raman, and X-Ray photoelectron spectroscopy. We demonstrated the method by evaluating the electrical performance of MoS$_2$-FETs and showed the effects of varying the sputtering power and duration. We found an optimum within our experimental split for a reverse sputtering power of 75 W for 10 s, as these devices exhibited a 50% reduction in $R_c$ compared to untreated reference devices. The reduced $R_c$ improved electrical performance metrics such as $I_{on}$, $I_{out}$, and mobility $\mu_{2p}$. The results can be explained by the introduction of sulfur vacancies by the sputtering process, which facilitates the phase transition from 2H- to 1T-MoS$_2$ and narrows the Schottky barrier width to the Ni contact metal. However, $R_c$ reduction was not achieved with inappropriate sputtering power, whether too weak (25 W) or too strong (200 W). Our findings suggest reverse sputtering as a highly efficient, large-scale, and CMOS-compatible method to optimize $R_c$ in MoS$_2$-FETs.

## Methods

Device fabrication: Si chips (2 × 2 cm$^2$) covered with 90 nm thermal SiO$_2$ were used as substrates. A multilayer MoS$_2$ material grown via metal–organic chemical vapor deposition (MOCVD) on sapphire developed by Grundmann *et al.* was used for this work.[43] The material was transferred onto the substrate via a wet transfer method with poly(methyl methacrylate) (PMMA) as a supporting layer. The PMMA/MoS$_2$ layer was delaminated from the as-grown sapphire wafer via a 4 mol potassium hydroxide (KOH) solution and cleaned from the etchant by floating on DI water overnight. TOP-FETs and BOT-FETs were fabricated via optical lithography (EVG 420 Mask Aligner) with an AZ5214E resist (Merck Performance Materials GmbH). In the case of bottom-contact FETs (BOT-FETs), a 50 nm thick layer of Ni was initially sputter-treated via direct current (DC) sputtering via the von Ardenne sputter tool "CS 730 S" onto the substrate. After the lift-off process of the Ni layer in 80 °C acetone, MoS$_2$ was transferred onto the substrate, and subsequent patterning of the MoS2 channel was achieved through reactive-ion etching (RIE) in an Oxford Instruments "Plasmalab System 100" tools with a mixture of tetrafluoromethane (CF$_4$) and oxygen gas. A reverse sputtering process was also achieved in the von Ardenne sputter tool in radio frequency (RF) mode, and the process pressure was 5 × 10$^{-3}$ Torr. The reverse sputtering process was performed on the channel material of the devices to investigate the electrical properties. On the other hand, the fabrication of TOP-FETs started from the wet transfer of MoS$_2$ onto the substrate. Reverse sputtering was applied after the RIE process for channel patterning, followed by Ni deposition of the top electrodes. Note that only the MoS$_2$ at the contact regions of the TOP-FETs was treated with reverse sputtering. In contrast, the MoS$_2$ in the channel region was protected by an AZ5214E photoresist during optical lithography, preserving its pristine state.

Material Characterization: Optical microscope images were recorded with a Leica INM100 microscope. Atomic force microscopy (AFM) measurements were performed on the as-grown MOCVD MoS$_2$ with an area of 5 µm × 10 µm on the sapphire wafer via Dimension Icon AFM by a Bruker instrument in tapping mode. Raman and PL measurements were conducted via an alpha300R WITec confocal Raman spectrometer with an excitation laser wavelength of 532 nm and a laser power of 1 mW. Measurements were performed in mapping mode at room temperature on an area of 5 × 5 µm with 2500 different points. The resolutions for PL and Raman were 300 g mm$^{-1}$ grating and 1800 g mm$^{-1}$ grating, respectively. The PL and Raman spectra were obtained by averaging the PL and Raman spectra measured at 2500 points. The film composition was analyzed via X-ray photoelectron spectroscopy (XPS) with a Thermo Scientific KA1066 spectrometer and monochromatic Al K-α X-rays at 1486.6 eV. To correct for potential charging effects in the XPS spectra, charge correction was applied to position the C−C bonding contribution from adventitious carbon at 284.8 eV. Contributions from O1s, S2p, Mo3d, and S1s (overlapping regions) were measured, and the atomic S to Mo ratios were calculated from the integrated peak areas via the appropriate sensitivity factors.

Electrical Measurements: The electrical characterization of the MoS$_2$-FETs consisted of measurements of the transfer and output characteristics. The measurements were performed in a cryogenic probe station "CRX-6.5K" from LakeShore Cryotronics connected to a Keithley

SCS4200 parameter analyzer. The probe station was conditioned either in vacuum ($3 \times 10^{-5}$ mBar) or under ambient conditions (42% humidity) in complete darkness and at room temperature (21 °C). Output curves were obtained by sweeping the $V_{DS}$ from 0 V to 4 V at $V_{GS}$ of 30 V to 40 V with an interval of 2 V, whereas transfer curves were obtained by sweeping the $V_{GS}$ from -40 V to 40 V at a $V_{DS}$ of 1 V.

# Acknowledgments

This work has received funding from the European Union's Horizon 2020 and Horizon Europe Research and Innovation Programme under grant agreements Nos. 881603 (GrapheneCore3), 863258 (ORIGENAL), 952792 (2D-EPL), 101189797 (2D-PL) and 101135571 (AttoSwitch), from the German Federal Ministry of Education and Research, BMBF, within the projects NEUROTEC (Nos. 16ES1134 and 16ES1133K), NEUROTEC 2 (Nos. 16ME0399, 16ME0398K, and 16ME0400), and NeuroSys (No. 03ZU1106AA and 03ZU1106BA) and from the German Research Foundation (DFG) under grants LE 2440/7-1, LE 2440/8-1 and WA 4139/3-1 is gratefully acknowledged.

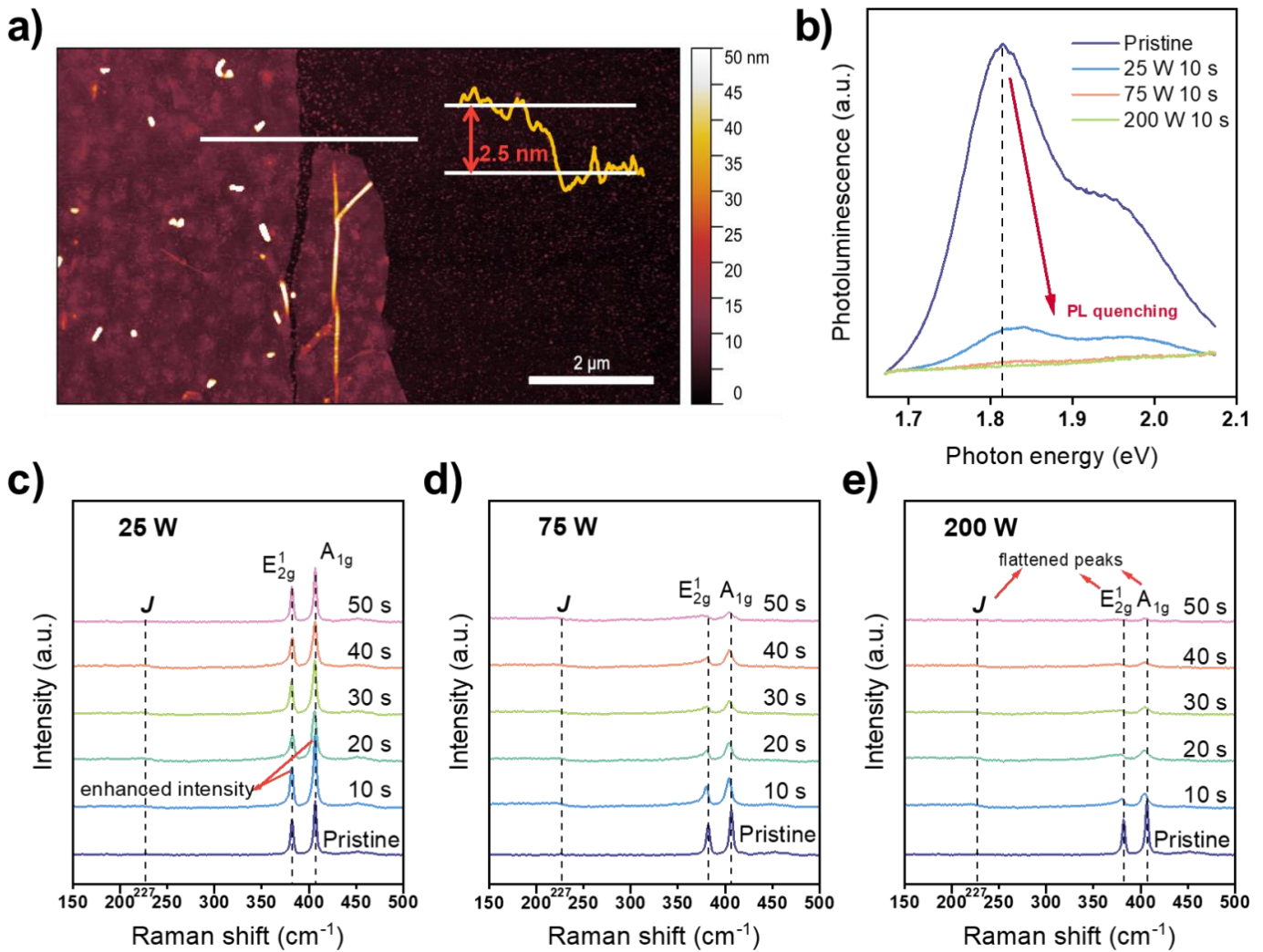

Figure 1: a) AFM image of MoS2 on a SiO2/Si substrate. The inset height profile shows a thickness of approximately 2.5 nm, indicating a multilayer MoS2 film. b) PL spectra of pristine $MoS_2$ and $MoS_2$ under reverse sputtering at various powers and durations, showing a trend of PL quenching. Raman spectra of pristine 2D $MoS_2$ and sputter-treated $MoS_2$ at powers of c) 25 W, d) 75 W, and e) 200 W with increasing treatment duration until 50 s. The spectra of pristine $MoS_2$ in purple are located at the bottom of each series. The emerging *J* peak indicates distorted octahedral coordination of $MoS_2$, and the gradually flat spectra at 75 W and 200 W suggest that the crystal structure of $MoS_2$ is gradually destroyed.

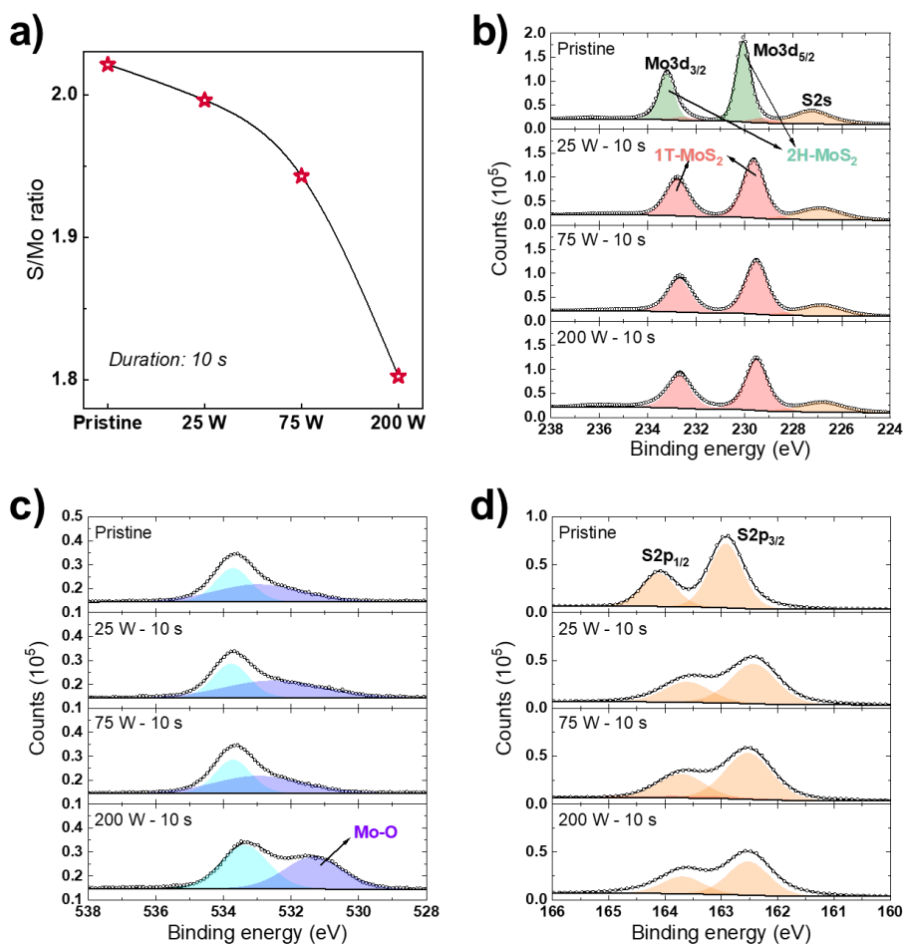

Figure 2: a) Sulfur to molybdenum ratio of the $MoS_2$ films as determined by XPS. The stoichiometry was calculated from the integrated peak areas of the Mo3d and S2p peaks. XPS spectra of b) Mo3d, c) S2p, and d) O1s. Note that there is an overlap with the S2s peak at a binding energy of approximately 227 eV. The spectra were fitted with three contributions for Mo3d. Two contributions corresponding to the 2H and 1T phases were fitted for the Mo3d$^{4+}$ state. For these two Mo3d$^{4+}$ contributions, the 1T phase contribution was constrained to have a 0.9 eV higher binding energy than the 2H contribution.[38] Upon sputtering, the O1s contribution at a lower binding energy increases.

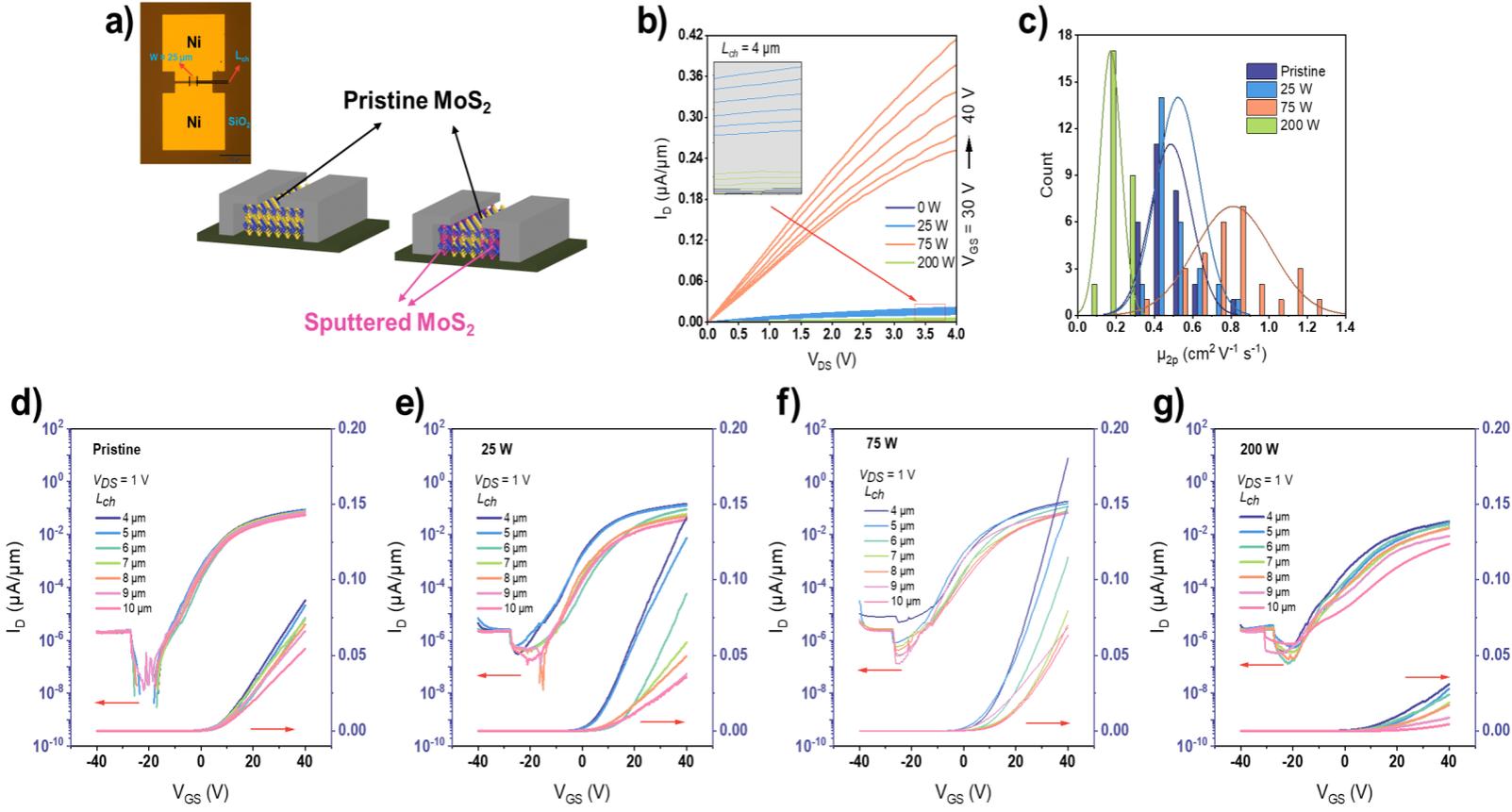

Figure 3: a) Schematics of a pristine MoS$_2$-FET (left) and a FET with a pristine MoS$_2$ channel combined with sputter-treated MoS$_2$ under the Ni electrodes (right). Inset: microscope image of a Ni-contacted MoS$_2$-FET. b) Output characteristics of TOP-FETs with L$_{ch}$ = 4 μm on the pristine sample (labeled 0 W) and samples under reverse sputtering of 25 W, 75 W, and 200 W. Inset: magnified view of the $I_D$ of the pristine sample and samples treated with 25 W and 200 W sputtering power. c) Histogram of $\mu_{2p}$ of 28 TOP-FETs on each sample. The mean value and standard deviation are extracted by fitting a Gaussian distribution to the histogram data, where 75 W shows an enhanced $\mu_{2p}$. d)-g) Transfer characteristics of pristine TOP-FETs and sputter-treated TOP-FETs at e) 25 W, f) 75 W, and g) 200 W in linear and logarithmic scales.

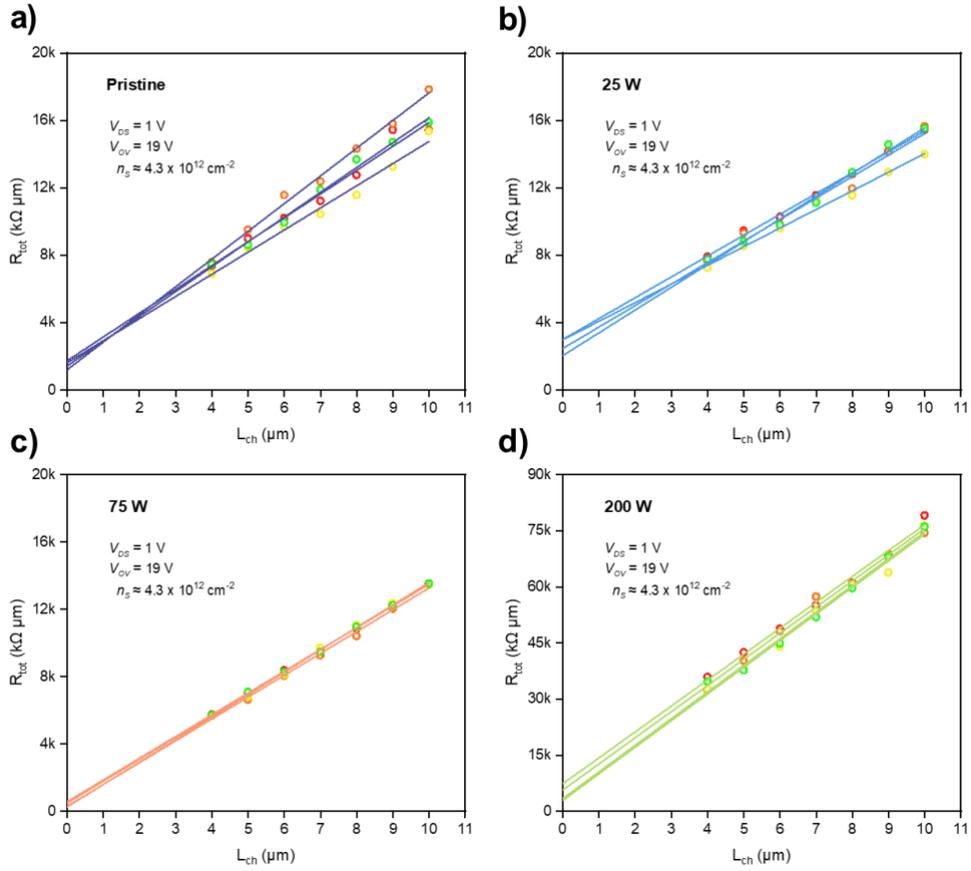

Figure 4: TLM plots of 4 groups of 7 TOP-FETs with L$_{ch}$ ranging from 4 to 10 μm on a) the pristine sample, b) 25 W, c) 75 W, and d) 200 W samples with the same $V_{OV}$ of 19 V.

Table 1: Mean $R_c$ and $R_{sh}$ with standard errors calculated via the TLM.

| Sample | $R_c$ (kΩ·μm) | $R_{sh}$ (kΩ/sq) |
|---|---|---|
| **Pristine** | 1126 ± 1053 | 3058 ± 289 |
| **25 W** | 2632 ± 412 | 2484 ± 113 |
| **75 W** | 413 ± 167 | 2613 ± 146 |
| **200 W** | 4883 ± 2377 | 14081 ± 653 |

# Reference

(1) Akinwande, D.; Huyghebaert, C.; Wang, C.-H.; Serna, M. I.; Goossens, S.; Li, L.-J.; Wong, H.-S. P.; Koppens, F. H. L. Graphene and Two-Dimensional Materials for Silicon Technology. *Nature* **2019**, *573* (7775), 507–518. https://doi.org/10.1038/s41586-019-1573-9.

(2) O'Brien, K. P.; Naylor, C. H.; Dorow, C.; Maxey, K.; Penumatcha, A. V.; Vyatskikh, A.; Zhong, T.; Kitamura, A.; Lee, S.; Rogan, C.; Mortelmans, W.; Kavrik, M. S.; Steinhardt, R.; Buragohain, P.; Dutta, S.; Tronic, T.; Clendenning, S.; Fischer, P.; Putna, E. S.; Radosavljevic, M.; Metz, M.; Avci, U. Process Integration and Future Outlook of 2D Transistors. *Nat. Commun.* **2023**, *14* (1), 6400. https://doi.org/10.1038/s41467-023-41779-5.

(3) Lembke, D.; Bertolazzi, S.; Kis, A. Single-Layer $MoS_2$ Electronics. *Acc. Chem. Res.* **2015**, *48* (1), 100–110. https://doi.org/10.1021/ar500274q.

(4) Kappera, R.; Voiry, D.; Yalcin, S. E.; Branch, B.; Gupta, G.; Mohite, A. D.; Chhowalla, M. Phase-Engineered Low-Resistance Contacts for Ultrathin MoS2 Transistors. *Nat. Mater.* **2014**, *13* (12), 1128–1134. https://doi.org/10.1038/nmat4080.

(5) Gan, X.; Lee, L. Y. S.; Wong, K.; Lo, T. W.; Ho, K. H.; Lei, D. Y.; Zhao, H. 2H/1T Phase Transition of Multilayer $MoS_2$ by Electrochemical Incorporation of S Vacancies. *ACS Appl. Energy Mater.* **2018**, *1* (9), 4754–4765. https://doi.org/10.1021/acsaem.8b00875.

(6) Schneider, D. S.; Lucchesi, L.; Reato, E.; Wang, Z.; Piacentini, A.; Bolten, J.; Marian, D.; Marin, E. G.; Radenovic, A.; Wang, Z.; Fiori, G.; Kis, A.; Iannaccone, G.; Neumaier, D.; Lemme, M. C. CVD Graphene Contacts for Lateral Heterostructure MoS2 Field Effect Transistors. *Npj 2D Mater. Appl.* **2024**, *8* (1), 35. https://doi.org/10.1038/s41699-024-00471-y.

(7) Zhu, J.; Wang, Z.; Yu, H.; Li, N.; Zhang, J.; Meng, J.; Liao, M.; Zhao, J.; Lu, X.; Du, L.; Yang, R.; Shi, D.; Jiang, Y.; Zhang, G. Argon Plasma Induced Phase Transition in Monolayer $MoS_2$. *J. Am. Chem. Soc.* **2017**, *139* (30), 10216–10219. https://doi.org/10.1021/jacs.7b05765.

(8) Xiao, J.; Chen, K.; Zhang, X.; Liu, X.; Yu, H.; Gao, L.; Hong, M.; Gu, L.; Zhang, Z.; Zhang, Y. Approaching Ohmic Contacts for Ideal Monolayer $MoS_2$ Transistors Through Sulfur-Vacancy Engineering. *Small Methods* **2023**, 2300611. https://doi.org/10.1002/smtd.202300611.

(9) Yoon, Y.; Ganapathi, K.; Salahuddin, S. How Good Can Monolayer $MoS_2$ Transistors Be? *Nano Lett.* **2011**, *11* (9), 3768–3773. https://doi.org/10.1021/nl2018178.

(10) Piacentini, A.; Marian, D.; Schneider, D. S.; González Marín, E.; Wang, Z.; Otto, M.; Canto, B.; Radenovic, A.; Kis, A.; Fiori, G.; Lemme, M. C.; Neumaier, D. Stable Al2O3 Encapsulation of MoS2-FETs Enabled by CVD Grown h-BN. *Adv. Electron. Mater.* **2022**, *8* (9), 2200123. https://doi.org/10.1002/aelm.202200123.

(11) Radisavljevic, B.; Radenovic, A.; Brivio, J.; Giacometti, V.; Kis, A. Single-Layer MoS2 Transistors. *Nat. Nanotechnol.* **2011**, *6* (3), 147–150. https://doi.org/10.1038/nnano.2010.279.

(12) Valasa, S.; Tayal, S.; Thoutam, L. R.; Ajayan, J.; Bhattacharya, S. A Critical Review on Performance, Reliability, and Fabrication Challenges in Nanosheet FET for Future

Analog/Digital IC Applications. *Micro Nanostructures* **2022**, *170*, 207374. https://doi.org/10.1016/j.micrna.2022.207374.

(13) Kumari, N. A.; Sreenivasulu, V. B.; Vijayvargiya, V.; Upadhyay, A. K.; Ajayan, J.; Uma, M. Performance Comparison of Nanosheet FET, CombFET, and TreeFET: Device and Circuit Perspective. *IEEE Access* **2024**, *12*, 9563–9571. https://doi.org/10.1109/ACCESS.2024.3352642.

(14) Uchida, K.; Koga, J.; Takagi, S. Experimental Study on Carrier Transport Mechanisms in Double- and Single-Gate Ultrathin-Body MOSFETs - Coulomb Scattering, Volume Inversion, and /Spl Delta/T/Sub SOI/-Induced Scattering. In *Electron Devices Meeting, 2003. IEDM '03 Technical Digest. IEEE International*; 2003; p 33.5.1-33.5.4. https://doi.org/10.1109/IEDM.2003.1269402.

(15) Schmidt, M.; Lemme, M. C.; Gottlob, H. D. B.; Driussi, F.; Selmi, L.; Kurz, H. Mobility Extraction in SOI MOSFETs with Sub 1 Nm Body Thickness. *Solid-State Electron.* **2009**, *53*, 1246–1251. https://doi.org/10.1016/j.sse.2009.09.017.

(16) Ellis, J. K.; Lucero, M. J.; Scuseria, G. E. The Indirect to Direct Band Gap Transition in Multilayered $MoS_2$ as Predicted by Screened Hybrid Density Functional Theory. *Appl. Phys. Lett.* **2011**, *99* (26), 261908. https://doi.org/10.1063/1.3672219.

(17) Wang, L.; Liu, X.; Luo, J.; Duan, X.; Crittenden, J.; Liu, C.; Zhang, S.; Pei, Y.; Zeng, Y.; Duan, X. Self-Optimization of the Active Site of Molybdenum Disulfide by an Irreversible Phase Transition during Photocatalytic Hydrogen Evolution. *Angew. Chem.* **2017**, *129* (26), 7718–7722. https://doi.org/10.1002/ange.201703066.

(18) Gao, G.; Jiao, Y.; Ma, F.; Jiao, Y.; Waclawik, E.; Du, A. Charge Mediated Semiconducting-to-Metallic Phase Transition in Molybdenum Disulfide Monolayer and Hydrogen Evolution Reaction in New 1T' Phase. *J. Phys. Chem. C* **2015**, *119* (23), 13124–13128. https://doi.org/10.1021/acs.jpcc.5b04658.

(19) Lin, Y.-C.; Dumcenco, D. O.; Huang, Y.-S.; Suenaga, K. Atomic Mechanism of the Semiconducting-to-Metallic Phase Transition in Single-Layered MoS2. *Nat. Nanotechnol.* **2014**, *9* (5), 391–396. https://doi.org/10.1038/nnano.2014.64.

(20) Piacentini, A.; Daus, A.; Wang, Z.; Lemme, M. C.; Neumaier, D. Potential of Transition Metal Dichalcogenide Transistors for Flexible Electronics Applications. *Adv. Electron. Mater.* **2023**, *9* (8), 2300181. https://doi.org/10.1002/aelm.202300181.

(21) Prakash, A.; Ilatikhameneh, H.; Wu, P.; Appenzeller, J. Understanding Contact Gating in Schottky Barrier Transistors from 2D Channels. *Sci. Rep.* **2017**, *7* (1), 12596. https://doi.org/10.1038/s41598-017-12816-3.

(22) Schwarz, M.; Vethaak, T. D.; Derycke, V.; Francheteau, A.; Iniguez, B.; Kataria, S.; Kloes, A.; Lefloch, F.; Lemme, M.; Snyder, J. P.; Weber, W. M.; Calvet, L. E. The Schottky Barrier Transistor in Emerging Electronic Devices. *Nanotechnology* **2023**, *34* (35), 352002. https://doi.org/10.1088/1361-6528/acd05f.

(23) Shen, P.-C.; Su, C.; Lin, Y.; Chou, A.-S.; Cheng, C.-C.; Park, J.-H.; Chiu, M.-H.; Lu, A.-Y.; Tang, H.-L.; Tavakoli, M. M.; Pitner, G.; Ji, X.; Cai, Z.; Mao, N.; Wang, J.; Tung, V.; Li, J.; Bokor, J.; Zettl, A.; Wu, C.-I.; Palacios, T.; Li, L.-J.; Kong, J. Ultralow Contact Resistance between Semimetal and Monolayer Semiconductors. *Nature* **2021**, *593* (7858), 211–217. https://doi.org/10.1038/s41586-021-03472-9.

(24) Li, W.; Gong, X.; Yu, Z.; Ma, L.; Sun, W.; Gao, S.; Köroğlu, Ç.; Wang, W.; Liu, L.; Li, T.; Ning, H.; Fan, D.; Xu, Y.; Tu, X.; Xu, T.; Sun, L.; Wang, W.; Lu, J.; Ni, Z.; Li, J.; Duan, X.; Wang, P.; Nie, Y.; Qiu, H.; Shi, Y.; Pop, E.; Wang, J.; Wang, X. Approaching the Quantum Limit in Two-Dimensional Semiconductor Contacts. *Nature* **2023**, *613* (7943), 274–279. https://doi.org/10.1038/s41586-022-05431-4.

(25) Mondal, A.; Biswas, C.; Park, S.; Cha, W.; Kang, S.-H.; Yoon, M.; Choi, S. H.; Kim, K. K.; Lee, Y. H. Low Ohmic Contact Resistance and High on/off Ratio in Transition Metal Dichalcogenides Field-Effect Transistors via Residue-Free Transfer. *Nat. Nanotechnol.* **2024**, *19* (1), 34–43. https://doi.org/10.1038/s41565-023-01497-x.

(26) Sun, Z.; Kim, S. Y.; Cai, J.; Shen, J.; Lan, H.-Y.; Tan, Y.; Wang, X.; Shen, C.; Wang, H.; Chen, Z.; Wallace, R. M.; Appenzeller, J. Low Contact Resistance on Monolayer $MoS_2$ Field-Effect Transistors Achieved by CMOS-Compatible Metal Contacts. *ACS Nano* **2024**, *18* (33), 22444–22453. https://doi.org/10.1021/acsnano.4c07267.

(27) Mootheri, V.; Arutchelvan, G.; Banerjee, S.; Sutar, S.; Leonhardt, A.; Boulon, M.-E.; Huyghebaert, C.; Houssa, M.; Asselberghs, I.; Radu, I.; Heyns, M.; Lin, D. Graphene Based Van Der Waals Contacts on $MoS_2$ Field Effect Transistors. *2D Mater.* **2021**, *8* (1), 015003. https://doi.org/10.1088/2053-1583/abb959.

(28) Frindt, R. F. Single Crystals of MoS2 Several Molecular Layers Thick. *J. Appl. Phys.* **1966**, *37* (4), 1928–1929. https://doi.org/10.1063/1.1708627.

(29) McDonnell, S. J.; Wallace, R. M. Atomically-Thin Layered Films for Device Applications Based upon 2D TMDC Materials. *Thin Solid Films* **2016**, *616*, 482–501. https://doi.org/10.1016/j.tsf.2016.08.068.

(30) Kang, N.; Paudel, H. P.; Leuenberger, M. N.; Tetard, L.; Khondaker, S. I. Photoluminescence Quenching in Single-Layer $MoS_2$ via Oxygen Plasma Treatment. *J. Phys. Chem. C* **2014**, *118* (36), 21258–21263. https://doi.org/10.1021/jp506964m.

(31) Eda, G.; Yamaguchi, H.; Voiry, D.; Fujita, T.; Chen, M.; Chhowalla, M. Photoluminescence from Chemically Exfoliated $MoS_2$. *Nano Lett.* **2011**, *11* (12), 5111–5116. https://doi.org/10.1021/nl201874w.

(32) Nayak, A. P.; Pandey, T.; Voiry, D.; Liu, J.; Moran, S. T.; Sharma, A.; Tan, C.; Chen, C.-H.; Li, L.-J.; Chhowalla, M.; Lin, J.-F.; Singh, A. K.; Akinwande, D. Pressure-Dependent Optical and Vibrational Properties of Monolayer Molybdenum Disulfide. *Nano Lett.* **2015**, *15* (1), 346–353. https://doi.org/10.1021/nl5036397.

(33) Livneh, T.; Sterer, E. Resonant Raman Scattering at Exciton States Tuned by Pressure and Temperature in 2 H -MoS 2. *Phys. Rev. B* **2010**, *81* (19), 195209. https://doi.org/10.1103/PhysRevB.81.195209.

(34) Li, H.; Zhang, Q.; Yap, C. C. R.; Tay, B. K.; Edwin, T. H. T.; Olivier, A.; Baillargeat, D. From Bulk to Monolayer MoS2: Evolution of Raman Scattering. *Adv. Funct. Mater.* **2012**, *22* (7), 1385–1390. https://doi.org/10.1002/adfm.201102111.

(35) Najmaei, S.; Liu, Z.; Ajayan, P. M.; Lou, J. Thermal Effects on the Characteristic Raman Spectrum of Molybdenum Disulfide (MoS2) of Varying Thicknesses. *Appl. Phys. Lett.* **2012**, *100* (1), 013106. https://doi.org/10.1063/1.3673907.

(36) Li, H.; Tsai, C.; Koh, A. L.; Cai, L.; Contryman, A. W.; Fragapane, A. H.; Zhao, J.; Han, H. S.; Manoharan, H. C.; Abild-Pedersen, F.; Nørskov, J. K.; Zheng, X. Activating and Optimizing MoS2 Basal Planes for Hydrogen Evolution through the Formation of Strained Sulphur Vacancies. *Nat. Mater.* **2016**, *15* (1), 48–53. https://doi.org/10.1038/nmat4465.

(37) Dieterle, M.; Mestl, G. Raman Spectroscopy of Molybdenum Oxides. *Phys. Chem. Chem. Phys.* **2002**, *4* (5), 822–826. https://doi.org/10.1039/b107046k.

(38) Lin, Z.; Liu, Y.; Halim, U.; Ding, M.; Liu, Y.; Wang, Y.; Jia, C.; Chen, P.; Duan, X.; Wang, C.; Song, F.; Li, M.; Wan, C.; Huang, Y.; Duan, X. Solution-Processable 2D Semiconductors for High-Performance Large-Area Electronics. *Nature* **2018**, *562* (7726), 254–258. https://doi.org/10.1038/s41586-018-0574-4.

(39) English, C. D.; Shine, G.; Dorgan, V. E.; Saraswat, K. C.; Pop, E. Improved Contacts to $MoS_2$ Transistors by Ultra-High Vacuum Metal Deposition. *Nano Lett.* **2016**, *16* (6), 3824–3830. https://doi.org/10.1021/acs.nanolett.6b01309.

(40) Li, Y.; Wang, Y.; Lin, J.; Shi, Y.; Zhu, K.; Xing, Y.; Li, X.; Jia, Y.; Zhang, X. Solution-Plasma-Induced Oxygen Vacancy Enhances MoOx/Pt Electrocatalytic Counter Electrode for Bifacial Dye-Sensitized Solar Cells. *Front. Energy Res.* **2022**, *10*, 924515. https://doi.org/10.3389/fenrg.2022.924515.

(41) Nazir, G.; Khan, M. F.; Iermolenko, V. M.; Eom, J. Two- and Four-Probe Field-Effect and Hall Mobilities in Transition Metal Dichalcogenide Field-Effect Transistors. *RSC Adv.* **2016**, *6* (65), 60787–60793. https://doi.org/10.1039/C6RA14638D.

(42) Islam, M. R.; Kang, N.; Bhanu, U.; Paudel, H. P.; Erementchouk, M.; Tetard, L.; Leuenberger, M. N.; Khondaker, S. I. Tuning the Electrical Property *via* Defect Engineering of Single Layer $MoS_2$ by Oxygen Plasma. *Nanoscale* **2014**, *6* (17), 10033–10039. https://doi.org/10.1039/C4NR02142H.

(43) Grundmann, A.; McAleese, C.; Conran, B. R.; Pakes, A.; Andrzejewski, D.; Kümmell, T.; Bacher, G.; Teo, K. B. K.; Heuken, M.; Kalisch, H.; Vescan, A. MOVPE of Large-Scale MoS2/WS2, WS2/MoS2, WS2/Graphene and MoS2/Graphene 2D-2D Heterostructures for Optoelectronic Applications. *MRS Adv.* **2020**, *5* (31–32), 1625–1633. https://doi.org/10.1557/adv.2020.104.

# Supporting Information

## Contact Resistance Optimization in MoS$_2$ Field-Effect Transistors through Reverse Sputtering-Induced Structural Modifications

Yuan Fa[1,2], Agata Piacentini[1,2], Bart Macco[3], Holger Kalisch[4], Michael Heuken[4,5], Andrei Vescan[4], Zhenxing Wang[1,*], and Max C. Lemme[1,2,*]

1 AMO GmbH, Advanced Microelectronic Center Aachen, Otto-Blumenthal-Str. 25, 52074 Aachen, Germany

2 Chair of Electronic Devices, RWTH Aachen University, Otto-Blumenthal-Str. 25, 52074 Aachen, Germany

3 Department of Applied Physics and Science Education, Eindhoven University of Technology, 5600 MB Eindhoven, The Netherlands

4 Compound Semiconductor Technology, RWTH Aachen University, 52074 Aachen, Germany

5 AIXTRON SE, 52134 Herzogenrath, Germany

Bottom-contacted Field-Effect Transistors (BOT-FETs)

BOT-FETs were fabricated as a control experiment, where the entire MoS$_2$ in the channel and contact regions was exposed to the sputter treatment. Electrical measurements were conducted on the BOT-FETs under ambient conditions to assess the electrical properties of the sputter-treated MoS$_2$ channels. The device schematic is displayed in Fig. S5a. The treatment and measurement sequence followed the protocol used during Raman spectroscopy. Reverse sputtering was applied incrementally in 10-second intervals up to 50 s at 25 W, followed by an additional treatment at 200 W for up to 30 s. In Fig. S5b, the $R_{tot}$, which includes the resistance of the two contact regions and the channel region, is plotted against the duration of reverse sputtering of 25 W for three BOT-FETs with L$_{ch}$ of 8, 9, and 10 μm and a W of 25 μm. Most strikingly, the $R_{tot}$ of the three BOT-FETs experienced an evident decrease after an initial increase of 10 s. Subsequent application of 200 W after 25 W triggered a decrease in $R_{tot}$ and then stabilized at a specific resistive level. In addition, compared with the overall decrease in $R_{tot}$, an abnormal increase in $R_{tot}$ may be attributed to the worse contact properties under reverse sputtering. After applying 200 W for 30 s, $R_{tot}$ decreased by two orders of magnitude compared with that of pristine BOT-FETs. We can infer that the MoS$_2$ crystalline structure is more sensitive to the sputtering power than to the sputtering duration. Fig. S5c shows the transfer curves of the BOT-FET with an L$_{ch}$ of 8 μm obtained via the same sputter treatment method as in Fig. S5b. The transfer curves for reverse sputtering at 75 W and 200 W are shown in Fig. S5d-e. The pristine BOT-FET exhibits gate modulation with an on/off ratio of 10$^4$. However, the gate modulation is severely weakened after treatment at 25 W for 10 s. The remaining transfer curves appear flat, suggesting the loss of gate modulation. Compared with the pristine sample, the $I_D$ of the BOT-FETs treated by 200 W reverse sputtering increased by more than five orders of magnitude, revealing the metallic property of the sputter-treated $MoS_2$. We also observed that the $I_D$ after 50 s of sputter treatment is slightly lower than after 40 s. This could be explained by the fact that the sputter-treated $MoS_2$ may be oxidized in the ambient environment during the measurement, which is in agreement with the XPS results.

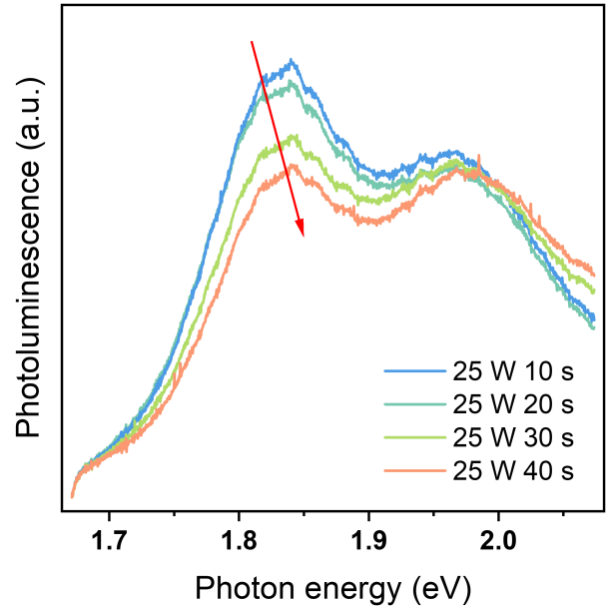

Figure S1: PL spectra at reverse sputtering at 25 W, showing gradually quenched PL peak as sputter duration increased.

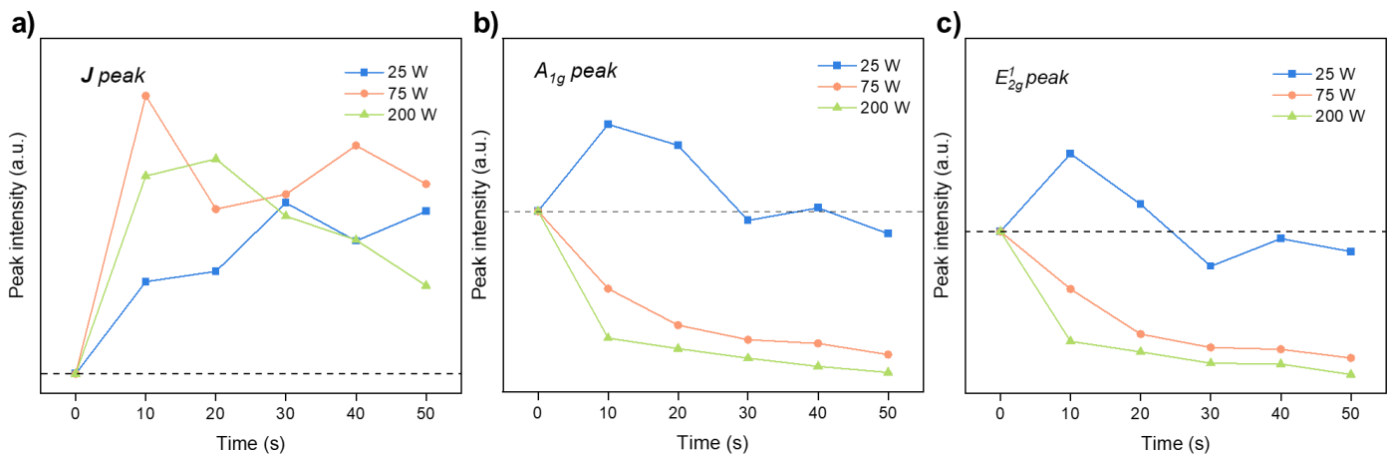

Figure S2: Changes in peak intensities of the a) *J* peak, b) $A_{1g}$ peak, c) and $E_{2g}^1$ peak with increasing duration of reverse sputtering at 25 W, 75 W, and 200 W.

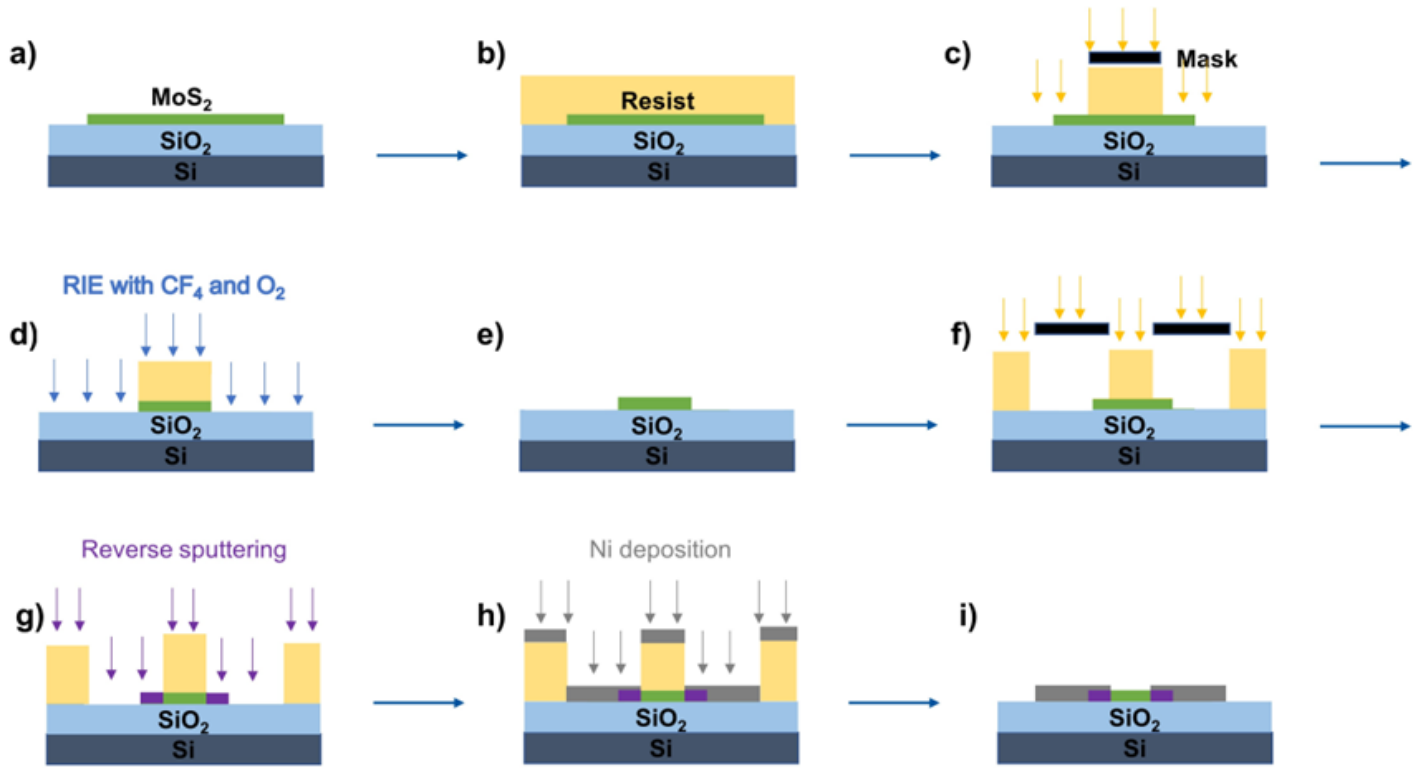

Figure S3: Schematic of the fabrication process of TOP-FETs with pristine channel MoS$_2$ and sputter-treated MoS2 as contact electrodes: (a) wet transfer of 2D MoS$_2$ onto the SiO$_2$/Si substrate, b) spin-coating of the AZ-5214E photoresist, c) positive lithography, d) RIE process, e) image reversal lithography, f) image reversal lithography, g) reverse sputtering, h) Ni sputtering process, and i) cleaning of the resist and the final TOP-FETs.

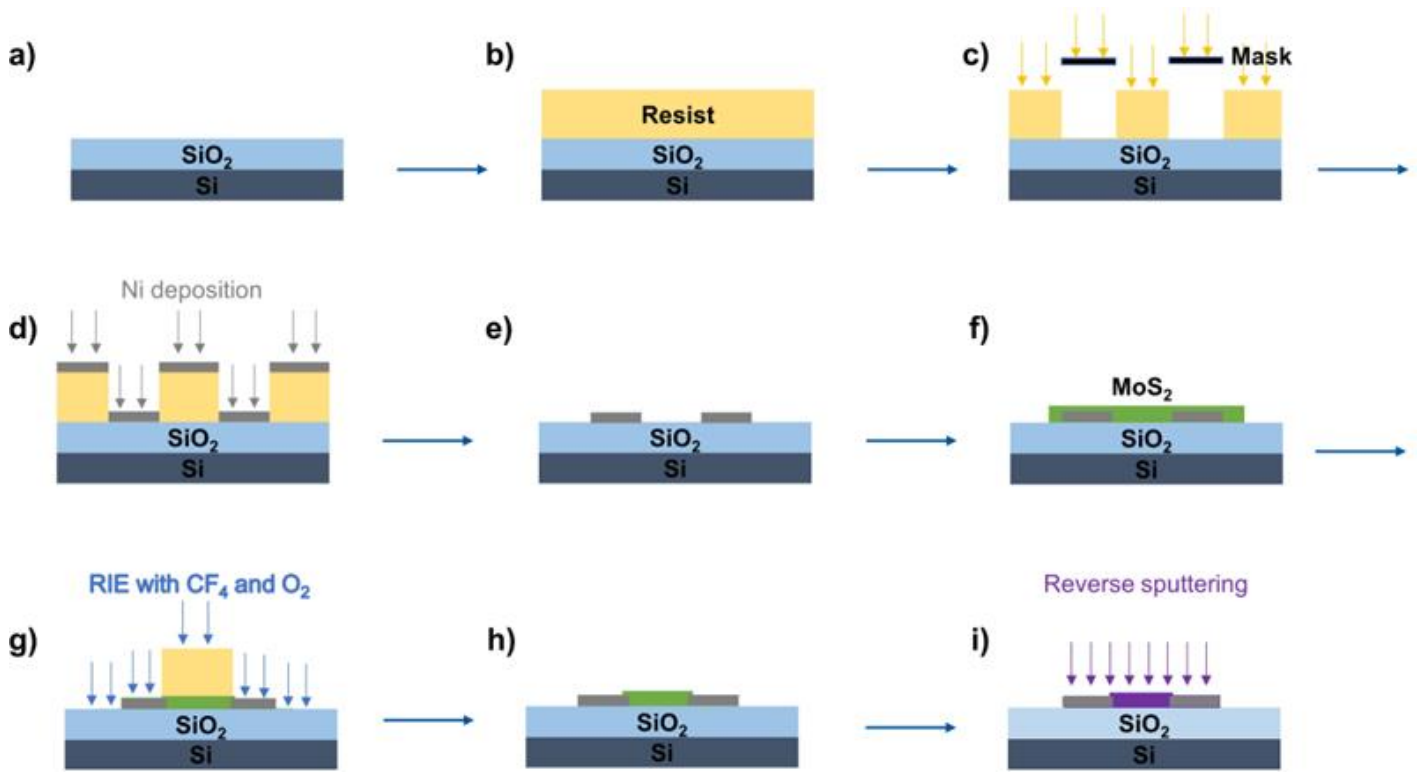

Figure S4: Schematic of the fabrication process of BOT-FETs with sputter-treated MoS$_2$: (a) Clean SiO$_2$/Si substrate, b) spin-coating of the AZ5214E resist, c) image reversal lithography, d) Ni sputtering process, e) lift-off process, f) wet transfer of 2D MoS$_2$ onto the metal pads, g) positive lithography and RIE process, h) cleaning of the photoresist, and i) reverse sputtering.

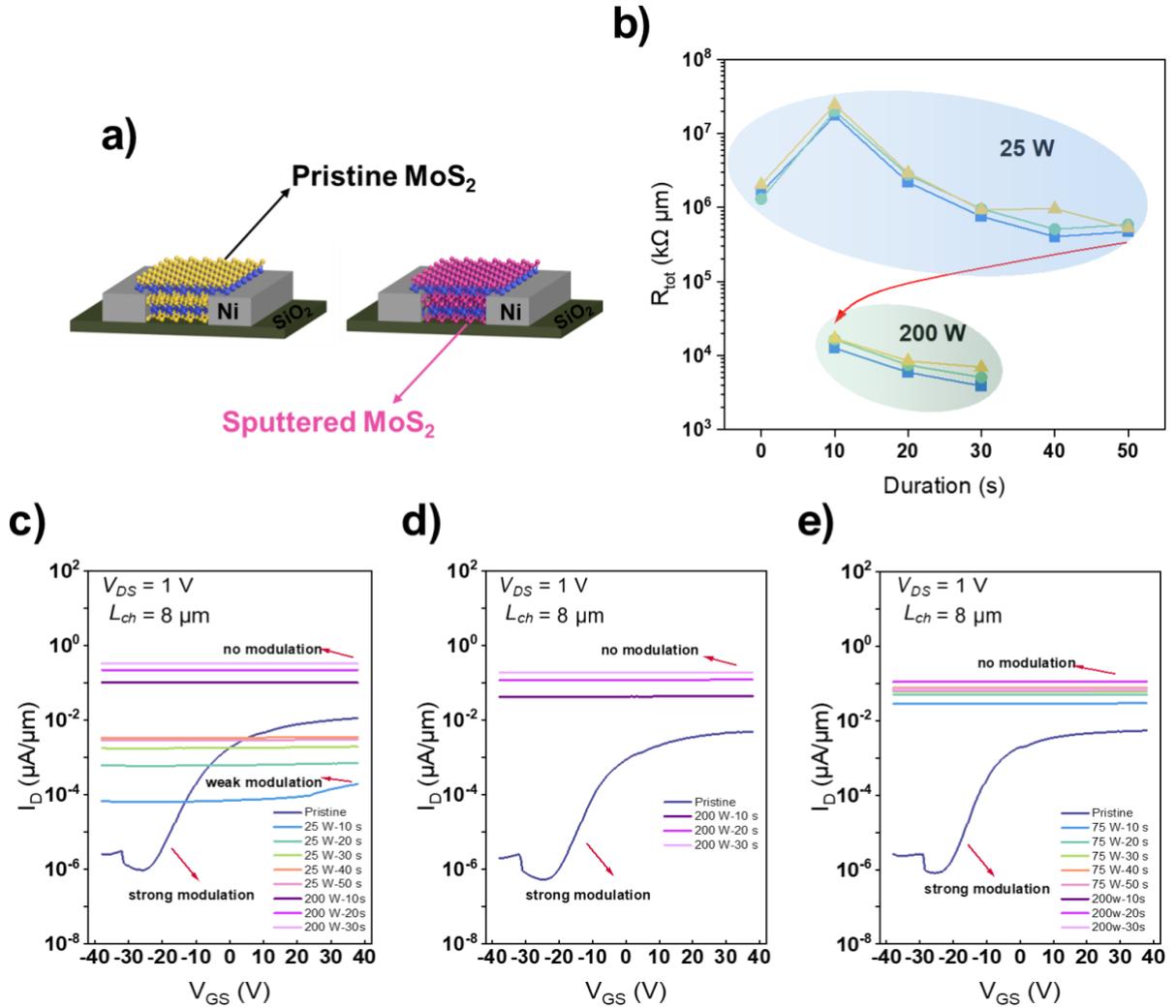

Figure S5: a) Schematics of the BOT-FETs. The MoS$_2$ in the right schematic is reverse sputter-treated with increasing treatment duration of up to 50 s. b) $R_{tot}$ versus time with 25 W sputtering (up to 50 s) and 200 W sputtering, showing an abrupt change in $R_{tot}$ and a weakened decreasing trend. c) Transfer curves of the BOT-FETs with the same treatment method as in b). The FET with pristine MoS$_2$ shows gate modulation, which is gradually lost after the FET is treated with reverse sputtering. The transfer curves of BOT-FETs subjected to reverse sputtering at d) 75 W and e) 200 W.